\documentclass{article}

\usepackage{amsmath}
\usepackage{amsthm}
\usepackage{amssymb}
\usepackage{amsfonts}

\newtheorem{theorem}{Theorem}[section]
\newtheorem{lemma}[theorem]{Lemma}
\newtheorem{corollary}[theorem]{Corollary}

\newtheorem{definition}[theorem]{Definition}

\newcommand{\E}{{\bf E}}
\newcommand{\RR}{\mathbb R}

\begin{document}

\title{A note on concentration of submodular functions}

\author{Jan Vondr\'ak\thanks{IBM Almaden Research Center, San Jose, CA 95120. E-mail: {\tt jvondrak@ibm.us.com}}}

\maketitle

\begin{abstract}
We survey a few concentration inequalities for submodular and fractionally subadditive
functions of independent random variables, implied by the entropy method for self-bounding functions.
The power of these concentration bounds is that they are dimension-free, in particular implying
standard deviation $O(\sqrt{\E[f]})$ rather than $O(\sqrt{n})$ which can be obtained
for any $1$-Lipschitz function of $n$ variables.
\end{abstract}

\section{Introduction}
\label{sec:intro}

In this note, we survey several concentration bounds for submodular and fractionally subadditive
functions of independent random variables.
These bounds are obtained by the {\em entropy method} for self-bounding functions
 \cite{BLM00,BLM03,MR06,BLM09}. This is a powerful technique developed over the last decade,
which in particular recovers Talagrand's inequality \cite{BLM09}. We also recommend
the lecture notes by G\'abor Lugosi \cite{Lugosi}.
The connection between self-bounding and submodular functions is quite simple
but perhaps not widely known. To our knowledge, the first application of self-bounding functions
in computer science appeared in \cite{HKLR05}. Similar concentration bounds for submodular functions
have been proved
recently by two sets of authors \cite{BH09,CV09}, unaware of the connection with self-bounding functions.
Hence this note, which might be useful in applications involving submodular functions.
Let us start with the definitions.

\begin{definition}
A set function $f:2^N \rightarrow \RR$ is
\begin{itemize}
 \item monotone, if $f(A) \leq f(B)$ for all $A \subseteq B \subseteq N$.
 \item submodular, if $f(A \cup B) + f(A \cap B) \leq f(A) + f(B)$ for all $A,B \subseteq N$.
 \item fractionally subadditive, if $f(A) \leq \sum \beta_i f(B_i)$ whenever $\beta_i \geq 0$
 and $\sum_{i:a \in B_i} \beta_i \geq 1 \ \forall a \in A$.
 \item subadditive, if $f(A \cup B) \leq f(A) + f(B)$ for all $A \subseteq B \subseteq N$.
\end{itemize}
\end{definition}

Observe that the definition of fractional subadditivity implies that
$f(\emptyset)=0$ (by taking $A = B_1 = \emptyset$ and $\beta_1=0$ or $\beta_1=2$).
It also implies monotonicity (by taking $A \subseteq B_1$ and $\beta_1=1$),
and hence nonnegativity.
The definition of subadditivity implies nonnegativity (by taking $A=B$), but not monotonicity.
Submodularity implies neither non-negativity nor monotonicity.
The property of being submodular is relevant for non-monotone functions
(the cut function in a graph is an example).

We also use the notions of marginal values and Lipschitz functions.

\begin{definition}
For a function $f:2^N \rightarrow \RR$, the marginal value of $j \in N$ with respect to $A \subset N$
is $f_A(j) = f(A \cup \{j\}) - f(A)$. $f$ is called $c$-Lipschitz, if its marginal values
are bounded by $c$ in absolute value.
\end{definition}

A function is monotone if and only if its marginal values are always non-negative.
Submodularity can be expressed equivalently by saying that marginal values $f_A(j)$
are non-increasing with respect to $A$.
Furthermore, the following relationships are known \cite{LLN06}.

\begin{lemma}
\label{lem:inclusion}
If $f$ is non-negative monotone submodular, then it is fractionally subadditive.
If $f$ is fractionally subadditive, then it is also subadditive.
\end{lemma}

These inclusions are strict, and there are simple examples separating the three classes \cite{Feige06}.
Consider $f:2^{[3]} \rightarrow \RR_+$ such that $f(\emptyset) = 0$ and $f(S) = 1$ whenever $|S|=1$ or $2$.
Then if $f$ is submodular, we must have $f([3]) \leq 1$. If we define $f([3]) = 3/2$, $f$ is not submodular
but it is fractionally subadditive. If we define $f([3]) = 2$, $f$ is not fractionally subadditive
but it is still subadditive. Defining $f([3]) > 2$ would not make the function even subadditive.

\section{Self-bounding functions}
\label{sec:self-bound}

{\em Self-bounding} functions were introduced by Boucheron, Lugosi and Massart
\cite{BLM00}. Self-bounding functions are defined more generally on product spaces;
here we restrict our attention to the hypercube $\{0,1\}^n$. We identify functions on $\{0,1\}^n$
with set functions on $N = [n]$ in a natural way.

\begin{definition}
A function $f:\{0,1\}^n \rightarrow \RR$ is self-bounding, if there are functions
$f_i:\{0,1\}^{n-1} \rightarrow \RR$ such that if we denote
$x^{(i)} = (x_1,\ldots,x_{i-1},x_{i+1},\ldots,x_n)$,
then for all $x$ and $i$,
$$ 0 \leq f(x) - f_i(x^{(i)}) \leq 1 $$
and
$$ \sum_{i=1}^{n} (f(x) - f_i(x^{(i)})) \leq f(x).$$
\end{definition}

A typical choice of $f_i$ is $f_i(x^{(i)}) = \min_{x_i} f(x)$, which for monotone functions
means $f_i(x^{(i)}) =  f(x_1,\ldots,x_{i-1},0,x_{i+1},\ldots,x_n)$.
First, we show that fractionally subadditive functions are self-bounding.
Hence, every non-negative monotone submodular function is also self-bounding.

\begin{lemma}
\label{lem:fract-subadd}
Every fractionally subadditive function $f:2^N \rightarrow \RR_+$
with marginal values in $[0,1]$ is self-bounding.
\end{lemma}

\begin{proof}
We use $f_i(x^{(i)}) = f(x_1,\ldots,x_{i-1},0,x_{i+1},\ldots,x_n)$, which means that
$f(x) - f_i(x^{(i)})$ is the marginal value of $i$ with respect to $x^{(i)}$.
Thus the condition $0 \leq f(x) - f_i(x^{(i)}) \leq 1$ is satisfied by assumption.

For a given $x \in \{0,1\}^n$, define $A = \{j: x_j=1\}$, $B_i = A \setminus \{i\}$
and $\beta_i = \frac{1}{n-1}$. Slightly abusing notation, we have $f(x) = f(A)$ and
$f_i(x^{(i)}) = f(B_i)$. We also have $\sum_{i:j \in B_i} \beta_i = (n-1) \frac{1}{n-1} = 1$
for each $j \in A$. Therefore, the definition of fractional subadditivity implies that
$$ f(x) = f(A) \leq \sum_{i=1}^{n} \beta_i f(B_i) = \frac{1}{n-1} \sum_{i=1}^{n} f_i(x^{(i)}).$$
This proves the condition $ \sum_{i=1}^{n} (f(x) - f_i(x^{(i)})) \leq f(x).$
\end{proof}

McDiarmid and Reed \cite{MR06} further refined the notion of self-bounding functions as follows.

\begin{definition}
A function $f:\{0,1\}^n \rightarrow \RR$ is $(a,b)$-self-bounding, if there are $a,b \geq 0$ and
functions $f_i:\{0,1\}^{n-1} \rightarrow \RR$ such that if we denote
$x^{(i)} = (x_1,\ldots,x_{i-1},x_{i+1},\ldots,x_n)$,
then for all $x$ and $i$,
$$ 0 \leq f(x) - f_i(x^{(i)}) \leq 1 $$
and
$$ \sum_{i=1}^{n} (f(x) - f_i(x^{(i)})) \leq a f(x) + b.$$
\end{definition}

We remark that subadditive functions are not always self-bounding, or even $(a,b)$-self-bounding
for any constant $a,b$. (See the example at the end of Section~\ref{sec:concentration}.)
However, the notion of $(a,b)$-self-bounding functions is useful for us, because {\em non-monotone submodular}
functions turn out to be $(2,0)$-self-bounding.

\begin{lemma}
\label{lem:nonmon-submod}
Every $1$-Lipschitz submodular function $f:2^N \rightarrow \RR_+$ is $(2,0)$-self-bounding.
\end{lemma}

\begin{proof}
We consider $f$ as a function on $\{0,1\}^n$, and define $f_i(x^{(i)}) = \min_{x_i} f(x)$.
Note that here, it is not always the case that $f_i(x^{(i)})$ is obtained by setting $x_i = 0$.
Denote by $A$ the indices $i$ where the minimum is attained for $x_i=0$,
and by $B$ the indices $i$ where the minimum is attained for $x_i=1$.
(In case of equality, say we assign the index to $A$.)
In both cases, we have $0 \leq f(x) - f_i(x^{(i)}) \leq 1$ since the marginal values
of $f$ are bounded by $1$ in absolute value.

We bound the sum of the marginal values in two steps. First, let us add up over all indices
in $A = \{a_1,\ldots,a_k\}$.
We denote by $x^0(A')$ the point where the coordinates on $A'$ have been set to zero.
By submodularity, we have
\begin{eqnarray*}
\sum_{i \in A} (f(x) - f_i(x^{(i)})) & = & \sum_{j=1}^{k} (f(x) - f(x^0(\{a_j\}))) \\
 & \leq & \sum_{j=1}^{k} (f(x^0(\{a_1,\ldots,a_{j-1}\})) - f(x^0(\{a_1,\ldots,a_j\}))) \\
 & = & f(x) - f(x^0(A)) \leq f(x)
\end{eqnarray*}
using $f(x^0(A)) \geq 0$. Similarly, we add up over the indices $B = \{b_1,\ldots,b_\ell\}$
where the minimum is attained by setting $x_i = 1$.
We denote by $x^1(B')$ the point where the coordinates on $B'$ have been set to $1$.
By submodularity, we have
\begin{eqnarray*}
\sum_{i \in B} (f(x) - f_i(x^{(i)})) & = & \sum_{j=1}^{\ell} (f(x) - f(x^1(\{b_j\}))) \\
 & \leq & \sum_{j=1}^{\ell} (f(x^1(\{b_1,\ldots,b_{j-1}\})) - f(x^1(\{b_1,\ldots,b_j\}))) \\
 & = & f(x) - f(x^1(B)) \leq f(x).
\end{eqnarray*}
Since $(A,B)$ is a partition of $[n]$, we conclude:
$$ \sum_{i=1}^{n} (f(x) - f_i(x^{(i)})) = \sum_{i \in A} (f(x) - f_i(x^{(i)}))
 + \sum_{i \in B} (f(x) - f_i(x^{(i)})) \leq 2 f(x).$$
\end{proof}

We remark that the factor of $2$ is necessary. For example, if $n=2$ and $f(x_1,x_2) = x_1 (1-x_2)$
(the cut function of one directed edge), we have $f(1,0) = 1$ but flipping each coordinate
decreases the value by $1$, which means $ \sum_{i=1}^{n} (f(x) - f_i(x^{(i)})) = 2$.

\section{Concentration of submodular and fractionally subadditive functions}
\label{sec:concentration}

Boucheron, Lugosi and Massart proved that self-bounding functions of independent
random variables are strongly concentrated, using the entropy method \cite{BLM00}.
They proved the following bound on the exponential moment of a self-bounding function.

\begin{theorem}
\label{thm:self-bounding}
If $Z = f(X_1,\ldots,X_n)$ where $X_i \in \{0,1\}$ are independently random and $f$ is self-bounding, then
for any $\lambda \in \RR$
$$ \log \E[e^{\lambda (Z - \E[Z])}] \leq (e^\lambda - \lambda - 1) \E[Z].$$
\end{theorem}

By Lemma~\ref{lem:fract-subadd} and Lemma~\ref{lem:inclusion}, this bound also holds for
any non-negative monotone submodular or fractionally subadditive function with marginal values in $[0,1]$.
Positive and negative choices of $\lambda$ yield Chernoff-type bounds for the upper and lower tails.
We pick $\lambda = \ln (1+\delta)$ for the upper tail, and $\lambda = \ln (1-\delta)$ for the lower tail.
Applying Markov's inequality to the exponential moment, we obtain tail estimates similar
to Chernoff-Hoeffding bounds.

\begin{corollary}
\label{cor:monotone-tails}
If $Z = f(X_1,\ldots,X_n)$ where $X_i \in \{0,1\}$ are independently random and $f$ is self-bounding
(or in particular, non-negative submodular or fractionally subadditive with marginal values in $[0,1]$),
then for any $\delta > 0$,
\begin{itemize}
\item
$\Pr[Z \geq (1+\delta) \E[Z]] \leq \left( \frac{e^\delta}{(1+\delta)^{1+\delta}} \right)^{\E[Z]}.$
\item
$ \Pr[Z \leq (1-\delta) \E[Z]]  \leq e^{- \delta^2 \E[Z] / 2}.$
\end{itemize}
\end{corollary}

The power of these concentration bounds is that they are {\em dimension-free}, i.e. independent of $n$.
In particular, they imply that $Z$ is concentrated around $\E[Z]$ with
standard deviation $O(\sqrt{\E[Z]})$. Weaker bounds with standard deviation
$O(\sqrt{n})$ can be obtained by martingale arguments for {\em any} $1$-Lipschitz function.

We remark that the tail estimates are often presented in a somewhat different form.
In particular, \cite{BLM09} presents the upper-tail bound as follows:
$\Pr[Z \geq \E[Z] + t] \leq e^{-t^2 / (2\E[Z] + 2t/3)}$.
This bound is easier to work with, but it becomes weaker for large $t$.
In particular, if $\E[Z]=1$, then it would seem that we need $t = \Omega(\log n)$ to make the probability
polynomially small in $n$, while Corollary~\ref{cor:monotone-tails} implies that
$t = \delta = \Omega(\log n / \log \log n)$ is sufficient. The difference can be crucial in some applications.

For non-monotone submodular functions, we need to use more general bounds
for $(a,b)$-self-bounding functions, which were proved in \cite{MR06} and
strengthened in \cite{BLM09}.

\begin{theorem}
\label{thm:(a,b-self-bounding}
If $Z = f(X_1,\ldots,X_n)$ where $X_i \in \{0,1\}$ are independently random and $f$ is $(a,b)$-self-bounding,
$a \geq 1/3$ and $c = (3a-1)/6$, then
\begin{itemize}
\item for any $t>0$, $\Pr[Z \geq \E[Z] + t] \leq e^{-\frac12 t^2 / (a \E[Z] + b + ct)}$,
\item for any $0<t\le\E[Z]$, $\Pr[Z \leq \E[Z] - t]  \leq e^{-\frac12 t^2 / (a \E[Z] + b)}$.
\end{itemize}
\end{theorem}

Since non-monotone submodular functions are $(2,0)$-self-bounding (Lemma~\ref{lem:nonmon-submod}),
we use this bound with $a=2$ and $b=0$, i.e. $c = 5/6$. We also substitute $t = \delta \, \E[Z]$.

\begin{corollary}
\label{cor:nonmonotone-tails}
If $Z = f(X_1,\ldots,X_n)$ where $X_i \in \{0,1\}$ are independently random and $f$ is non-negative
submodular with marginal values in $[-1,1]$, then for any $\delta>0$,
\begin{itemize}
\item $\Pr[Z \geq (1+\delta) \E[Z]] \leq e^{-\delta^2 \E[Z] / (4 + 5\delta/3)}.$
\item $\Pr[Z \leq (1-\delta) \E[Z]] \leq e^{-\delta^2 \E[Z] / 4}.$
\end{itemize}
\end{corollary}

Observe that here, the upper tail decays only as a simple exponential for $\delta \rightarrow \infty$,
i.e. it is slightly weaker than the Chernoff-type bound in Corollary~\ref{cor:monotone-tails}.
We do not know whether this is necessary.

\section{Subadditive functions}
\label{sec:subadditive}

Finally, we show that subadditive functions are not $(a,b)$-self-bounding for any fixed $a,b>0$ and in fact
do not enjoy concentration properties similar to Corollaries~\ref{cor:monotone-tails}, \ref{cor:nonmonotone-tails}.

\medskip
\noindent
{\bf A counterexample.}
\begin{itemize}
 \item $f(S) = |S|$ for $|S| < \sqrt{n}$,
 \item $f(S) = \sqrt{n}$ for $\sqrt{n} \leq |S| \leq (n - \sqrt{n})/2$,
 \item $f(S) = \sqrt{n} + |S| - (n - \sqrt{n})/2$ for $(n-\sqrt{n})/2 < |S| < (n + \sqrt{n})/2$,
 \item $f(S) = 2\sqrt{n}$ for $|S| \geq (n + \sqrt{n})/2$.
\end{itemize}
Clearly, the marginal values are in $[0,1]$. We claim that this function is subadditive.
Consider $f(A) + f(B)$: If both $|A|,|B| \geq \sqrt{n}$, then we have $f(A)+f(B) \geq 2\sqrt{n} \geq f(A \cup B)$.
So assume $|A| < \sqrt{n}$. Then $f(A) + f(B) = |A| + f(B) \geq f(A \cup B)$, because the marginal values
are at most $1$ and hence $f(A \cup B) - f(B)$ cannot be more than $|A|$.

This function is not $(a,b)$-self-bounding for any constant $a,b$: $f(S) = \frac32 \sqrt{n}$ for $|S| = n/2$
(assuming $n$ even), and we get $f(S) - f(S \setminus \{i\}) = 1$ for all $i \in S$.
Therefore, $\sum_{i=1}^{n} (f(S) - f(S \setminus \{i\})) = n/2$.

Indeed, this function is not sharply concentrated. Consider a uniformly random set $R$
(corresponding to independent random unbiased variables $X_i \in \{0,1\}$). We have $|R| \leq (n-\sqrt{n})/2$
with constant probability (roughly $1/\sqrt{2 \pi e} \simeq 0.24$, from the central limit theorem),
and the same holds for $|R| \geq (n+\sqrt{n})/2$. Therefore, with constant probabilities, $f(R)$ is either
$\sqrt{n}$ or $2\sqrt{n}$, and the expectation is roughly $\frac32 \sqrt{n}$. The standard
deviation is on the order of $\sqrt{n}$ - such a bound can be obtained for any $1$-Lipschitz function.

Still, subadditive functions satisfy some concentration properties that do not hold
for arbitrary $1$-Lipschitz functions. In particular, it is very improbable
that a subadditive function attains a value significantly above {\em 3 times its median}.
The following more general inequality is shown in \cite{Schechtman} (see Corollary 12, which
applies more generally to product spaces).

\begin{theorem}
\label{thm:subadditive}
Let $Z = f(X_1,\ldots,X_n)$ where $f$ is non-negative subadditive with marginal values in $[0,1]$,
and $X_i \in \{0,1\}$ are independently random.
Then for any $a>0, 1 \leq k \leq n$ and integer $q \geq 18$,
$$ \Pr[Z \geq (q+1)a + k] \leq \Pr[Z \leq a]^{-q} q^{-k}.$$
\end{theorem}

In particular, if $a$ is the median of $Z$ and $q=2$, we get
$ \Pr[Z \geq 3a + k] \leq 2^{2-k}.$
Of course this is not true for an arbitrary non-negative monotone $1$-Lipschitz function.
Let for example $f(S) = \max \{0, |S|-n/2\}$, and $Z = f(R)$ where $R$ is uniformly random.
Then the median of $Z$ is $0$, but $Z \geq \sqrt{n}$ with constant probability.

\paragraph{Acknowledgment.}
We are indebted to Mohammad Taghi Hajiaghayi for pointing out
the connection which prompted us to write up this short survey.
Thanks are also due to Chandra Chekuri, Rico Zenklusen and Nick Harvey for useful comments.

\end{document}